\PassOptionsToPackage{svgnames, dvipsnames}{xcolor}

\documentclass[sigconf, screen]{acmart}

\usepackage{enumitem}

\usepackage{amssymb} 

\usepackage[utf8]{inputenc}

\usepackage{xspace}

\newcommand{\etal}{{~\textit{et al.}}}

\usepackage{tikz}
\usetikzlibrary{shadings}
\definecolor{cback}{HTML}{E9ECEF}
\definecolor{cframe}{HTML}{495057}

\newcommand*\circled[1]{\tikz[baseline=(char.base)]{
    \node[shape=rectangle,rounded corners=1.5pt,fill=cback,text=black,draw=cframe,inner sep=1pt] (char) {#1};}}

\newcommand{\ac}[1]{#1}
\newcommand{\ad}[1]{#1}
\newcommand{\rc}[1]{#1}
\newcommand{\rv}[1]{#1}

\newcommand{\cc}[1]{#1}

\usepackage[nameinlink,capitalise]{cleveref}

\AtBeginDocument{%
  }

\copyrightyear{2026}
\acmYear{2026}
\setcopyright{cc}
\setcctype{by}
\acmConference[SUI '26]{ACM Symposium on Spatial User Interaction}{October 10--11, 2026}{Bari, Italy}
\acmBooktitle{ACM Symposium on Spatial User Interaction (SUI '26), October 10--11, 2026, Bari, Italy}
\acmDOI{10.1145/3822518.3830041}
\acmISBN{979-8-4007-2812-9/2026/10}

\begin{document}

\title{Surrounded by Friends: Design and Evaluation of Immersive Layouts of Egocentric Network for Visual Analytics}

\author{Kentaro Takahira}
\orcid{0009-0003-5613-610X}
\affiliation{%
  \institution{HKUST}
  \city{Hong Kong}
  \country{China}
}
\email{ktakahira@connect.ust.hk}

\author{Takanori Fujiwara}
\orcid{0000-0002-6382-2752}
\affiliation{%
  \institution{University of Arizona}
  \city{Tucson}
  \state{Arizona}
  \country{United States}
}
\email{tfujiwara@arizona.edu}

\author{Wong Kam-Kwai}
\orcid{0000-0002-2813-1972}
\affiliation{%
  \institution{HKUST}
  \city{Hong Kong}
  \country{China}
}
\email{kkwongar@connect.ust.hk}

\author{Kento Shigyo}
\orcid{0000-0002-5095-7500}
\affiliation{%
  \institution{HKUST}
  \city{Hong Kong}
  \country{China}
}
\email{kshigyo@connect.ust.hk}

\author{Leni Yang}
\orcid{0000-0003-4527-4905}
\affiliation{%
  \institution{Inria, CNRS}
  \city{Bordeaux}
  \country{France}
}
\email{leni.yang@inria.fr}

\author{Hiroaki Natsukawa}
\orcid{0000-0001-6754-7834}
\affiliation{%
  \institution{Osaka Seikei University}
  \city{Osaka}
  \country{Japan}
}
\email{natsukawa@g.osaka-seikei.ac.jp}

\author{Yalong Yang}
\orcid{0000-0001-9414-9911}
\affiliation{%
  \institution{Georgia Institute of Technology}
  \city{Atlanta}
  \state{Georgia}
  \country{United States}
}
\email{yalong.yang@gatech.edu}

\author{Huamin Qu}
\orcid{0000-0002-3344-9694}
\affiliation{%
  \institution{HKUST}
  \city{Hong Kong}
  \country{China}
}
\email{huamin@cse.ust.hk}
\renewcommand{\shortauthors}{Takahira et al.}


\begin{abstract}
Egocentric networks focus on a focal node (\textit{ego}) and its neighboring \textit{alters}, emphasizing local sub-networks rather than the whole graph.
On traditional desktops, such visualizations suffer from clutter as networks grow, whereas immersive environments offer added spatial depth and embodied interaction. 
We explore how to design egocentric network layouts for these spaces.
We first identify essential design properties and dimensions tailored to immersive environments.
Based on these, we design four layouts--\textit{Cube}, \textit{Cylindrical}, \textit{Radial}, and \textit{Spherical}--that vary across design dimensions. 
We evaluate these layouts in a user study with 24 participants completing egocentric analysis tasks.
\rv{Our study suggests} that \textit{Cube} \rv{performed} well for tasks focused on ego-alter connection strength. 
In contrast, \textit{Spherical} \rv{was} more effective for understanding alter topology, minimizing occlusion, and efficiently utilizing 3D space. 
These findings inform the design of future immersive egocentric network layouts.
\end{abstract}




\begin{abstract}
Egocentric networks focus on a focal node (\textit{ego}) and its neighboring \textit{alters}, emphasizing local sub-networks rather than the whole graph.
On traditional desktops, such visualizations suffer from clutter as networks grow, whereas immersive environments offer added spatial depth and embodied interaction. 
We explore how to design egocentric network layouts for these spaces.
We first identify essential design properties and dimensions tailored to immersive environments.
Based on these, we design four layouts--\textit{Cube}, \textit{Cylindrical}, \textit{Radial}, and \textit{Spherical}--that vary across design dimensions. 
We evaluate these layouts in a user study with 24 participants completing egocentric analysis tasks.
\rv{Our study suggests} that \textit{Cube} \rv{performed} well for tasks focused on ego-alter connection strength. 
In contrast, \textit{Spherical} \rv{was} more effective for understanding alter topology, minimizing occlusion, and efficiently utilizing 3D space. 
These findings inform the design of future immersive egocentric network layouts.
\end{abstract}


\begin{teaserfigure}
  \centering
  \includegraphics[width=1\textwidth]{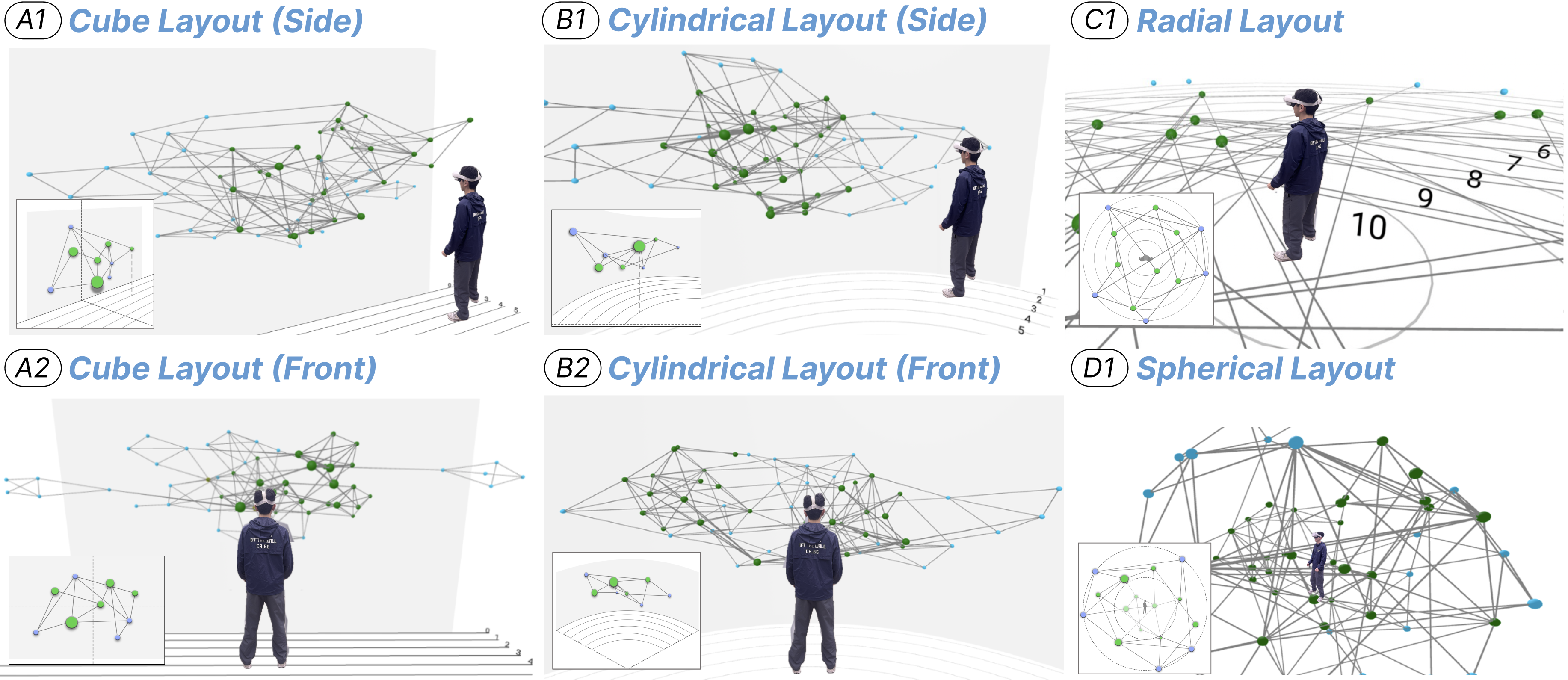}
  \caption{The Evaluated Immersive Egocentric Network Layouts:
(A) \textit{\textbf{Cube}}: Encodes ego–alter \cc{connection} strength using depth along the Z-axis.
(B) \textbf{\textit{Cylindrical}}: Uses radii of semi-cylinders centered on the user to represent \cc{ego-alter} connection strength.
(C) \textbf{\textit{Radial}}: Immersive extension of a radial layout common in desktop environments, centered at the user’s feet.
(D) \textbf{\textit{Spherical}}: Encodes \cc{ego-alter} connection strength via radii of spheres centered on the user.
}
\Description{XXX}
  \label{fig:teaser}
\end{teaserfigure}


\definecolor{clightblue}{HTML}{DDEFFC}
\definecolor{clightgreen}{HTML}{DFFFE5}
\definecolor{clightyellow}{HTML}{FFF4DB}

\definecolor{cback1}{HTML}{F8F9FA} 
\definecolor{cframe1}{HTML}{ADB5BD} 

\definecolor{cback2}{HTML}{F1F3F5} 
\definecolor{cframe2}{HTML}{CED4DA} 

\definecolor{cback3}{HTML}{EBF5FB} 
\definecolor{cframe3}{HTML}{A9CCE3} 

\definecolor{cback4}{HTML}{FDF3E7} 
\definecolor{cframe4}{HTML}{F5CBA7} 

\newcommand*\circledVarA[1]{\tikz[baseline=(char.base)]{
    \node[shape=rectangle,rounded corners=1.5pt,fill=cback1,text=black,draw=cframe1,inner sep=1pt] (char) {#1};}}

\newcommand*\circledVarB[1]{\tikz[baseline=(char.base)]{
    \node[shape=rectangle,rounded corners=1.5pt,fill=cback2,text=black,draw=cframe2,inner sep=1pt] (char) {#1};}}

\newcommand*\circledVarC[1]{\tikz[baseline=(char.base)]{
    \node[shape=rectangle,rounded corners=1.5pt,fill=cback3,text=black,draw=cframe3,inner sep=1pt] (char) {#1};}}

\newcommand*\circledVarD[1]{\tikz[baseline=(char.base)]{
    \node[shape=rectangle,rounded corners=1.5pt,fill=cback4,text=black,draw=cframe4,inner sep=1pt] (char) {#1};}}

\maketitle
\section{Introduction}
Egocentric networks \cc{focus on a central node} (\textit{ego}) and its connections to neighboring nodes (\textit{alters}). 
Egocentric analysis examines both the strengths of ego-alter connections and the topology among alters.
This approach is widely applied in fields such as sociology, bioinformatics, and public health~\cite{perry_egocentric_2018}, offering insights into personal relationships across life stages~\cite{degenne2005dynamics}, and behaviors shaped by network structure, such as health-related habits~\cite{russell2023social}.

However, visualizing egocentric networks is challenging as it requires simultaneous representation of \cc{ego-alter connection} strengths and alter topology. 
Radial layouts, which place the ego centrally and encode \cc{connection} strength by proximity, are commonly used~\cite{perry_egocentric_2018}.
Although intuitive and space-efficient~\cite{xue_target_2023}, these layouts suffer from visual clutter as network size grows~\cite{fu_dyegovis_2021,bauer_multi-layout_2023}. Despite various modifications to reduce clutter~\cite{xue_target_2023,brandes_more_nodate}, balancing clarity with connection-strength encoding remains challenging.
Alternative approaches using visual attributes such as color, edge thickness, and node size provide greater flexibility in node placement~\cite{wu_egoslider_2016, fu_dyegovis_2021,liu_egocomp_2017}. 
However, these channels are less effective than position, and the lack of a central reference weakens alter-hierarchy cues and spatial memory~\cite{brandes_communicating_2003,radialVisStrength}.

Virtual Reality (VR) offers new opportunities to overcome these limitations~\cite{kwon2016study, sorger_egocentric_2021, feyer_2d_2023}: stereoscopic depth conveys 3D structure, reducing clutter and improving spatial comprehension~\cite{ware_evaluating_1996, ware_visualizing_2008, munzner2008pitfalls}, \cc{while physical navigation and direct object manipulation~\cite{cordeil_design_nodate}} foster intuitive exploration~\cite{huang2017gesture}.
However, design strategies for effectively applying these capabilities to egocentric network analysis and how different layouts influence task performance remain underexplored.

To address this gap, we first identify two desirable properties of egocentric layouts: (1) clear depiction of ego-alter connection strength and (2) clear representation of \cc{alter-alter} topology. 
\cc{Building on these desirable properties and drawing from prior research on desktop-based egocentric layouts,} we identify four unique design dimensions: (1) positional encoding of ego-alter connection strength, (2) viewpoint assumptions, (3) spatial layout dimensions, and (4) the primary exploration strategies. 
Based on these dimensions, we propose four immersive layout designs---\textit{Cube}, \textit{Cylindrical}, \textit{Radial}, and \textit{Spherical}---each \cc{designed} to satisfy desirable properties and representing diverse design choices \ac{(\cref{fig:teaser})}. 

We evaluated these layouts with 24 participants performing representative egocentric \cc{network} tasks, examining each layout's strengths, limitations, and interaction patterns.
The results show diverse interaction patterns and task efficiencies across the layouts, with \textit{Cube} excelling in tasks focused on understanding ego-alter connection strengths and \textit{Spherical} proving more effective for alter topology comprehension. Additionally, we explore how different layouts and design dimensions impact users' task strategies. 
Based on these insights, we discuss the lessons learned and summarize design guidelines to enhance egocentric network analysis in immersive environments.
Our main contributions are:
\begin{itemize}[noitemsep,topsep=0pt,label=$\diamond$, leftmargin=*] 
\item Identification of key design dimensions for egocentric network visualization in immersive environments. 
\item Development of four novel layouts (\textit{Cube}, \textit{Cylindrical}, \textit{Radial}, and \textit{Spherical}) that span these dimensions.

\item A controlled user study evaluating the layouts and providing actionable design implications. \end{itemize}

\section{Related work}
\label{sec:related_work}

Our research builds on desktop-based egocentric network visualization \cc{as well as immersive network visualization and analytics.}

\subsection{Egocentric Network Analysis}
\label{sec:related_work_ego}

Egocentric networks focus on a central \textit{ego} and its directly connected \textit{1st alters}~\cite{perry_egocentric_2018, wu_egoslider_2016, liu_egocomp_2017}; analyses sometimes also include \textit{2nd alters}, connected to 1st alters but not to the ego~\cite{liu_egocomp_2017, EHLERS2024104123}.
Such analysis emphasizes ego–\cc{alter} relationships and interactions among alters, as individual behaviors are largely shaped by immediate relationships~\cite{perry_egocentric_2018, prell2011social, ProtEGOnist}.
Common analytical tasks include examining individual egocentric networks~\cite{perry_egocentric_2018, ProtEGOnist}, comparing multiple egocentric networks~\cite{liu_egocomp_2017, abbasi2012egocentric, ProtEGOnist}, and studying network dynamics over time~\cite{fu_dyegovis_2021, wu_egoslider_2016, shi_15d_2015, spreadline}.

It is widely applied across disciplines to reveal \cc{how network structures influence} individual and collective behaviors~\cite{provan_interorganizational_2007, prell2011social, fisher2005using, EHLERS2024104123}: in social network analysis to explore social support and information flow~\cite{degenne2005dynamics, ARNABOLDI201744, abbasi2012egocentric}; in epidemiology to model disease transmission~\cite{russell2023social}; and in organizational studies to examine communication and group dynamics~\cite{Chen_Gable_2013}.
Traditional visualizations often use node-link diagrams, most commonly radial layouts~\cite{fisher2005using, hollstein2020collecting, perry_egocentric_2018, EHLERS2024104123}, which position the ego centrally with alters in concentric circles by connection strength~\cite{perry_egocentric_2018, brandes_more_nodate, chung2005exploring}.
However, they grow cluttered with link crossings and node occlusion as network size grows~\cite{xue_target_2023, brandes_more_nodate}, impairing readability~\cite{fu_dyegovis_2021, EHLERS2024104123}. Solutions such as stress-based layouts~\cite{brandes_more_nodate} and annulus-constrained stress models~\cite{xue_target_2023} alleviate some occlusion and aid cluster detection but still struggle to encode strength precisely by distance, especially in dense networks.
Alternatives encode strength via color, edge thickness, and node size~\cite{wu_egoslider_2016, fu_dyegovis_2021, liu_egocomp_2017}, allowing more flexible placement, but these channels are less effective than position and lack a clear central reference, reducing salience and hindering recall of node positions and hierarchies~\cite{brandes_communicating_2003, radialVisStrength}.

A central challenge is balancing positional encoding with \rc{topological} clarity. Immersive environments, leveraging spatial dimensions and enhanced depth perception, present unique opportunities for improved egocentric network \cc{analysis}.

\subsection{Network Visualization in VR}
Immersive technologies open new possibilities for network analysis through 3D space and embodied interaction.
Early studies by Ware and Franck~\cite{ware_evaluating_1996}, and later by Ware and Mitchell~\cite{ware_visualizing_2008}, showed that stereoscopic displays with head-tracking---now standard in head-mounted displays---enhance depth perception and facilitate intuitive interpretation of 3D network structures.
Their 360-degree space also reduces node overlap and edge crossings common in 2D displays~\cite{kwon2016study}, while the added spatial dimension supports novel layouts~\cite{bauer_multi-layout_2023, sorokin_ring_2018, sorger_egocentric_2021, Sanky} and embodied navigation through physical movement and hand gestures~\cite{embodiedNetworkYalong}, helping users form clear mental models of network structure~\cite{kotlarek_study_2020}.

Recent studies have investigated diverse \cc{visual analytics} approaches for immersive networks~\cite{VRNetworkSurvey}. Kwon\etal~\cite{kwon2016study,7156357} examined spherical layouts and interaction methods, demonstrating their effectiveness for specific network tasks. Bauer\etal~\cite{bauer_multi-layout_2023} integrated 3D networks with 2D projections to provide additional contextual cues. Sorokin\etal~\cite{sorokin_ring_2018} arranged network nodes on spherical rings based on categorical attributes, effectively utilizing 3D space for detailed analyses. Other research has targeted particular structures, such as adjacency matrices for triangle detection~\cite{pan_extending_2023} and layered visualizations for multi-layer network tasks~\cite{feyer_2d_2023}. Additionally, various interaction techniques, including gesture-based node manipulation~\cite{huang2017gesture}, flying~\cite{drogemuller_evaluating_2018}, \cc{walking}~\cite{drogemuller2017vrige}, and teleporting to nodes~\cite{8942334}, have been introduced to enhance immersive navigation.

However, most immersive network visualization research emphasizes broad overviews rather than detailed subnetwork analysis. Sorger\etal~\cite{sorger_egocentric_2021} introduced an egocentric \cc{network} visualization \cc{in immersive environments}, allowing exploration from the perspective of selected nodes to examine local subnetworks. While this approach effectively reduces clutter and facilitates local topology comprehension, it lacks explicit encoding of ego-alter relationship strength, \cc{which is} crucial in egocentric analysis. \cc{Additionally, the effectiveness of this method} for specific egocentric network tasks remains largely untested.
\begin{figure*}[t!]
    \centering
    \includegraphics[width=1\linewidth]{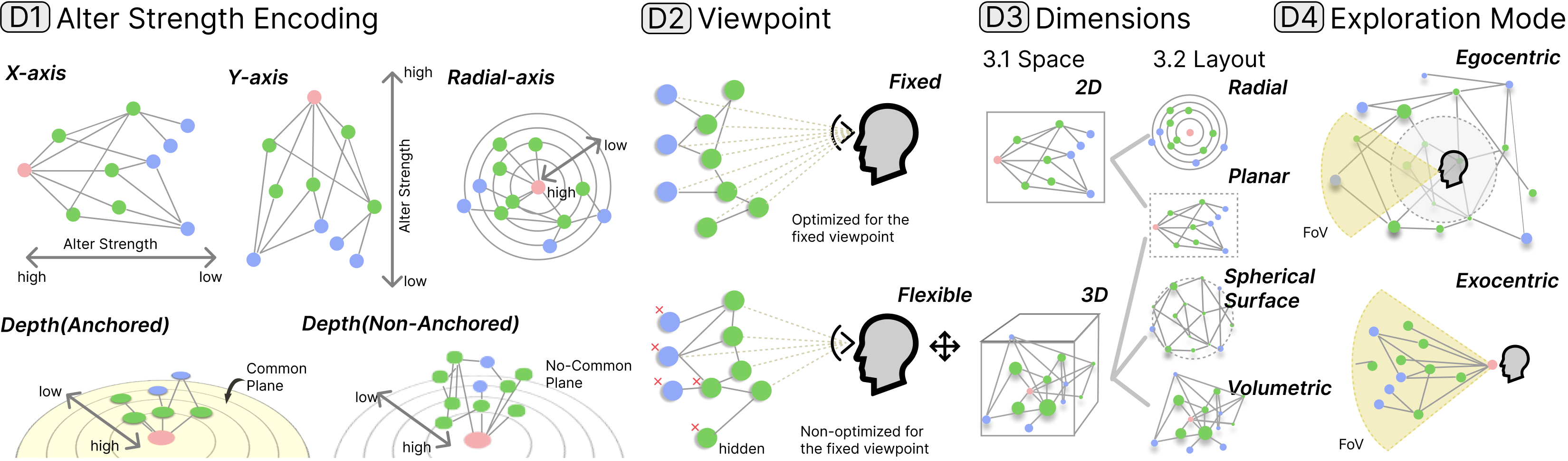}
    \caption{Design Dimensions for Egocentric Network Layouts in Immersive Environments: The red node represents the ego, green nodes are 1st-degree alters, and blue nodes are 2nd-degree alters.  
(D1) Alter strength can be encoded along the \( X/Y \) axes, radial axis, or depth, either with or without a shared reference plane.
(D2) Viewpoint assumptions vary from fixed to flexible.
(D3) Spatial layout dimensionality ranges from 2D to 3D.
(D4) Primary exploration mode can be either egocentric or exocentric.}
    \label{fig:designDimension}
\end{figure*}

\section{Egocentric Network Layouts}
\label{sec:design_and_layout}
This study explores layout designs for egocentric network analysis in \rc{VR} environments. \rv{We focus on a single static egocentric network and examine how layout design affects user performance.} We first define desirable properties and design dimensions, and then introduce four layout designs grounded in these considerations and prior research.

\subsection{Desirable Properties}
\label{sec:Desirable}

\subsubsection*{\circled{P1} \textbf{Effective Encoding of Ego–Alter Connection Strength}}
The ego-alter connection strength is central to egocentric network analysis~\cite{liu_egocomp_2017,perry_egocentric_2018}. 
While visual attributes of links (e.g., width, color) can encode this strength, they often contribute to visual clutter~\cite{perry_egocentric_2018}. 
Because every alter is directly connected to the ego, this strength can also be treated as a node attribute~\cite{liu_egocomp_2017, chung2005exploring}. 
\cc{Common approaches include encoding strength via node size, color, or position, which tend to be more scalable and visually clear}~\cite{bezerianos2010graphdice, sorokin_ring_2018}.

\subsubsection*{\circled{P2} \textbf{Effective Representation of Alter–Alter Topology:}}
Understanding the topology among 1st alters is another central goal~\cite{perry_egocentric_2018}. 
Analysts often investigate centrality, cluster formation, and the ego's structural role (e.g., as an articulation point between groups)~\cite{perry_egocentric_2018, ahuja2000collaboration, prell2011social}.
Additionally, analyzing both 1st and 2nd alters reveals extended structures and bridge roles, such as 1st alters connected to many 2nd alters, which may indicate influence or network reach~\cite{wu_egoslider_2016}.

\subsection{Design Dimensions}
Building on insights from desktop-based egocentric network layouts, we examine how these concepts extend to immersive environments. We identify four key design dimensions critical for egocentric layouts in VR: \circledVarA{D1} positional encoding of \cc{connection} strength, \circledVarA{D2} viewpoint assumptions, \circledVarA{D3} spatial layout dimensions, and \circledVarA{D4} primary exploration modes (\cref{fig:designDimension}).

\subsubsection{Positional Encoding of Alter Strength (D1)}
In desktop environments, ego-alter connection strength is often shown via node position~\cite{perry_egocentric_2018}.
In immersive environments, this positional encoding can be applied along horizontal, vertical, radial, or depth axes (\cref{fig:designDimension}-D1).
Horizontal and vertical encodings are generally effective, whereas radial encoding is less precise due to the difficulty of interpreting angular differences.
\cc{While depth-based encoding, which uses proximity to the ego to signal stronger connections, seems intuitive, users often struggle to perceive depth as accurately as horizontal or vertical positions~\cite{perceptionRanking}.}
Depth encoding can be anchored---nodes share a reference plane that aids depth perception---or non-anchored, distributing nodes \ac{freely in depth around the user, which hampers depth judgments but may enhance spatial immersion~\cite{slater2009place, egoVsExoYalong}.}

\subsubsection{Viewpoint Assumptions (D2)}
A key design consideration is whether the layout assumes a fixed viewpoint (\cref{fig:designDimension}-D2).
Layouts optimized for a specific viewpoint can reduce occlusion, improve initial comprehension, and help mitigate motion sickness~\cite{liu2020design}, but may restrict exploration from alternative angles. Conversely, viewpoint-independent layouts support multi-perspective exploration and deeper engagement, yet demand more movement and viewpoint adjustments, increasing cognitive load and motion sickness~\cite{9133071,10687394}, and may lack the clarity of fixed-viewpoint designs.

\subsubsection{Spatial Layout Dimensions (D3)}
The spatial dimensionality of a layout shapes both the user’s understanding of network structures and the flexibility of layout algorithms (\cref{fig:designDimension}-D3). 
2D layouts are intuitive for users familiar with desktop-based visualizations and, in immersive environments, can be adapted into forms such as curved layouts that wrap around the user or projections onto a virtual floor~\cite{liu2020design, datadancing}.
Such adaptations naturally guide user perspective and reduce cognitive load. 
\ac{Beyond the dimensionality of the space in which the network resides, each layout imposes distinct constraints on node placement.} 
\cc{In 2D space, radial layouts restrict nodes to concentric rings at fixed distances from the ego, limiting positional flexibility~\cite{brandes_more_nodate}.}
\ac{Planar layouts allow free placement on a flat surface, offering more flexibility than radial layouts, though they remain susceptible to link crossings due to the absence of vertical separation.}
\ac{In 3D space, spherical surface layouts constrain nodes to lie on the surface of a virtual sphere. In contrast, volumetric layouts allow nodes to be freely distributed throughout 3D space.}
\cc{These 3D layouts support natural user movement to resolve occlusion and enhance immersion~\cite{embodiedNetworkYalong, bethedata}. However, they can also increase cognitive load due to the challenges of navigating and interpreting complex 3D structures~\cite{feyer_2d_2023, breves2023cognitive}.}


\subsubsection{Primary Exploration Mode (D4)}
Layout design in immersive environments typically emphasizes either an exocentric view or an egocentric view as the primary exploration mode~\cite{sorger_egocentric_2021, effectOfExplorationMode}~(\cref{fig:designDimension}-D4). 
The exocentric view provides an external, global perspective, enabling users to grasp the overall structure at a glance. This viewpoint supports a strong sense of scale and helps users build an initial mental model, with the ability to zoom in for detailed inspection as needed. However, additional navigation may be required to examine specific subnetworks or node-level details closely.
In contrast, the egocentric view places the user inside the network, offering a first-person perspective that fills their field of vision. 
This immersive setup emphasizes subnetworks immediately surrounding the user, allowing focused inspection with minimal occlusion~\cite{sorger_egocentric_2021}. 
To explore the broader network, however, users must navigate across different areas. Prior work suggests that egocentric views can reduce cognitive load by enabling users to focus on one subregion at a time and improve memory recall through embodied spatial cues~\cite{effectOfExplorationMode}. Both views are widely used in VR-based network visualization~\cite{7156357, sorokin_ring_2018, sorger_egocentric_2021, 8942334}, each offering advantages depending on task and exploration phase.

\begin{figure*}[ht]
    \centering
    \includegraphics[width=1\linewidth]{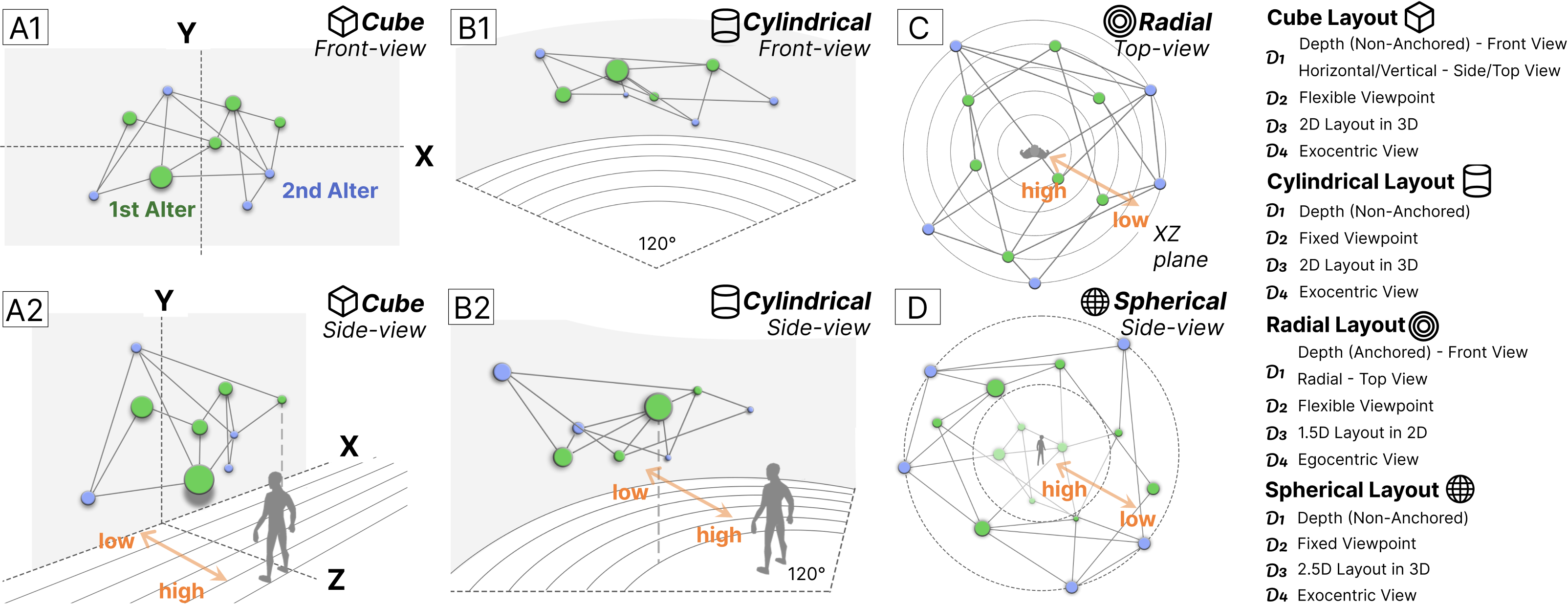}
    \caption{The Four Proposed Layouts:
    (A1) Front view of Cube Layout. (A2) Side view of Cube Layout. (B1) Front view of Cylindrical Layout. (B2) Side view of Cylindrical Layout. (C) Top view of Radial Layout. (D) Side view of Spherical Layout. Across all layouts, 1st alters are colored green, and 2nd alters are colored blue. The orange arrows denote the direction of the encoded value. The human figure illustrates the user's initial viewpoint in immersive environments.}
    \label{fig:layoutCompares}
\end{figure*}

\subsection{Immersive Layouts of Egocentric Network}
\label{sec:four_layouts}
Building on the desirable properties (\circled{P1}, \circled{P2}) and the design dimensions, \cc{we introduce four immersive layouts for egocentric networks---\textit{Cube}, \textit{Cylindrical}, \textit{Radial}, and \textit{Spherical}---for further exploration} (\cref{fig:layoutCompares}). 
\ac{In selecting and designing these layouts, we considered the unique potentials of immersive environments, particularly the use of spatial depth and flexible, viewpoint-driven navigation. 
These affordances create new opportunities to rethink egocentric network layouts beyond the limitations of traditional 2D representations. 
At the same time, we ensured that the selected layouts collectively span a broad range of combinations across the four design dimensions, enabling comparative analysis in the subsequent user study.}
Following conventions from prior visualizations such as target diagrams~\cite{TargetDiagram, brandes_communicating_2003}, we do not display the ego node or its \cc{links} to 1st alters, reducing visual clutter; first- and second-degree alters are instead distinguished visually. In the following layout descriptions, we define the floor as the \( XZ \)-plane and the vertical axis as \( Y \), assuming a standing user in an immersive space.

\subsubsection{Cube Layout}
The \textit{Cube} layout encodes ego–alter \cc{connection} strength along the \(Z\)-axis in 3D space. It begins with a 2D force-directed layout on the \( XY \)-plane (\cref{fig:layoutCompares}-A1), after which nodes are repositioned along the \( Z \)-axis based on their connection strength to the ego; stronger connections appear closer to the user's initial viewpoint (\cref{fig:layoutCompares}-A2).
Because nodes are distributed across multiple depth layers without a shared reference plane, users must interpret depth without anchored depth cues when viewing from the front.
\cc{The user is positioned outside the network, offering an exocentric perspective that reveals the overall structure.}
This layout encourages flexible exploration from multiple angles, such as side views (\( XZ \) or \( YZ \) planes), allowing users to interpret \cc{connection} strength through both horizontal and vertical positioning rather than relying solely on frontal depth cues.

\subsubsection{Cylindrical Layout}
The \textit{Cylindrical} layout is a curved variant of the \textit{Cube} layout, wrapped around the user along the \( Y \)-axis at their initial position (\cref{fig:layoutCompares}-B1).
It begins with a 2D force-directed layout on the \( XY \)-plane, which is then bent semi-cylindrically around the user. To encode ego–alter connection strength, each node's radial distance from the \( Y \)-axis is adjusted; nodes closer to the axis represent stronger ties (\cref{fig:layoutCompares}-B2). 
Unlike \textit{Cube}, the \textit{Cylindrical} layout is optimized for a fixed viewpoint, reducing occlusion by curving the network around the user. However, this limits side views, so users must rely primarily on frontal depth cues to assess connection strength.

\subsubsection{Radial Layout}
The \textit{Radial} layout adapts the widely used 2D radial layout for egocentric networks~\cite{brandes_communicating_2003, liu_egocomp_2017, TargetDiagram} to immersive environments (\cref{fig:layoutCompares}-C). 
Nodes are positioned on the floor (\( XZ \)-plane) using the radial layout algorithm by Brandes~\etal~\cite{brandes_communicating_2003}, with concentric circles centered on the user's initial position. Each node's radial distance from the center represents its connection strength to the ego. 
\cc{This layout restricts node positions to fixed distances from the center, which limits placement flexibility and can lead to occlusion or link crossings.} 
Because the network surrounds the user in 360 degrees, only a subnetwork is visible at a time. However, anchoring nodes to the floor provides a depth reference, aiding in depth perception. Users can also move vertically to obtain a top-down view, effectively replicating the traditional 2D radial layout.

\subsubsection{Spherical Layout}
The \textit{Spherical} layout arranges nodes using a spherical force-directed algorithm, positioning them within a 3D sphere centered on the user's initial location (\cref{fig:layoutCompares}-D).
Ego–alter \cc{connection} strength is encoded by each node's radial distance from the center, with stronger ties placed closer to the user. This design is inspired by the \textit{Ego-Bubble} layout proposed by Sorger\etal~\cite{sorger_egocentric_2021}, but differs in two key ways: the radial distance reflects connection strength, and a spherical force-directed algorithm is employed to preserve clustering patterns.
Unlike 2D force-directed layouts, the spherical algorithm allows greater flexibility in node placement and creates a fully immersive 360-degree network view. The layout is optimized for a fixed central viewpoint, minimizing occlusion from that position. 
Because the user is situated ``inside'' the network, they cannot view the entire structure at once.

\subsubsection{Other Layouts}
\cc{We conducted internal tests on various layout candidates and selected four---\textit{Cube}, \textit{Cylindrical}, \textit{Radial}, and \textit{Spherical}---for further evaluation.} 
\ac{These layouts were chosen for their ability to clearly expose trade-offs across the identified design dimensions and to effectively leverage key affordances of immersive environments. }
\cc{During the design exploration, we tested layouts that encoded ego–alter strength along the \(X\) or \(Y\) axes by first placing nodes on a plane using a 2D algorithm and then displacing them along one axis.}
\ac{However, these approaches resulted in severe visual clutter from the initial viewpoint and made poor use of the depth dimension.}
\cc{We also explored a cone-shaped variant of the radial layout, similar to cone trees~\cite{conetree}, which encoded strength in depth and helped reduce link crossings when users adjusted their viewpoint.}
\ac{Despite this benefit, it sacrificed the clarity provided by an anchored plane, making strength assessment more difficult. We ultimately selected the conventional \textit{Radial} layout instead, as it was the only candidate employing anchored depth encoding and served as a comparative baseline, reflecting the common approach for egocentric network visualization in desktop settings.}

\section{User Study}
\label{sec:user-study}
We conducted a user study to examine user performance across different \cc{immersive} egocentric network layouts. 
\rv{This study received approval from the Institutional Review Board at Hong Kong University of Science and Technology and proceeded only after the subjects signed consent forms.}

\subsection{Tasks}
Our user-study tasks are grouped by the focus of egocentric network analysis: Tasks 1 and 2 target ego–alter connection strength, Tasks 3 through 6 target alter topology, and Task 7 combines both dimensions, requiring participants to consider connection strength and topology simultaneously.
These tasks are grounded in previous research on egocentric networks~\cite{russell2023social, abbasi2012egocentric, perry_egocentric_2018, wu_egoslider_2016, fu_dyegovis_2021, sorger_egocentric_2021} and general network task taxonomies~\cite{networkTaskTaxonomy}, aligning with our desirable properties \circled{P1} (strength) and \circled{P2} (topology) described in Section~\ref{sec:Desirable}.

\subsubsection*{\circled{P1} \textbf{Task 1: Identify Alter with Highest Strength to Ego.}}
Participants select the 1st alter with the strongest connection to the ego, which is typically located near the initial viewpoint but not necessarily the closest. 

\subsubsection*{\circled{P1} \textbf{Task 2: Identify Alter with Lowest Strength to Ego.}}
From three randomly highlighted 1st alters, participants identify the one with the weakest connection to the ego (\cref{fig:screen-capture}-B). 
We limited the selection to three nodes based on pilot testing to ensure task feasibility. As weaker connections tend to lie farther from the participant, this task requires assessing distant nodes.

\subsubsection*{\circled{P2} \textbf{Task 3: Identify Alter with Most Neighboring Nodes.}}
Among the three highlighted nodes, participants select the one with the most connections to other alters, including both 1st and 2nd alters. 
This task reflects the identification of central nodes within the alter network; as in Task 2, the choice was limited to three nodes.

\subsubsection*{\circled{P2} \textbf{Task 4: Find Common Neighbor.}}
Participants identify a common neighbor between a randomly highlighted node pair. This task supports topology understanding of the alter network, requiring participants to either observe multiple nodes simultaneously or recall their connections. To ensure consistent difficulty, each highlighted pair shares between three and six common neighbors.

\begin{figure}[t]
    \centering
    \includegraphics[width=1\linewidth]{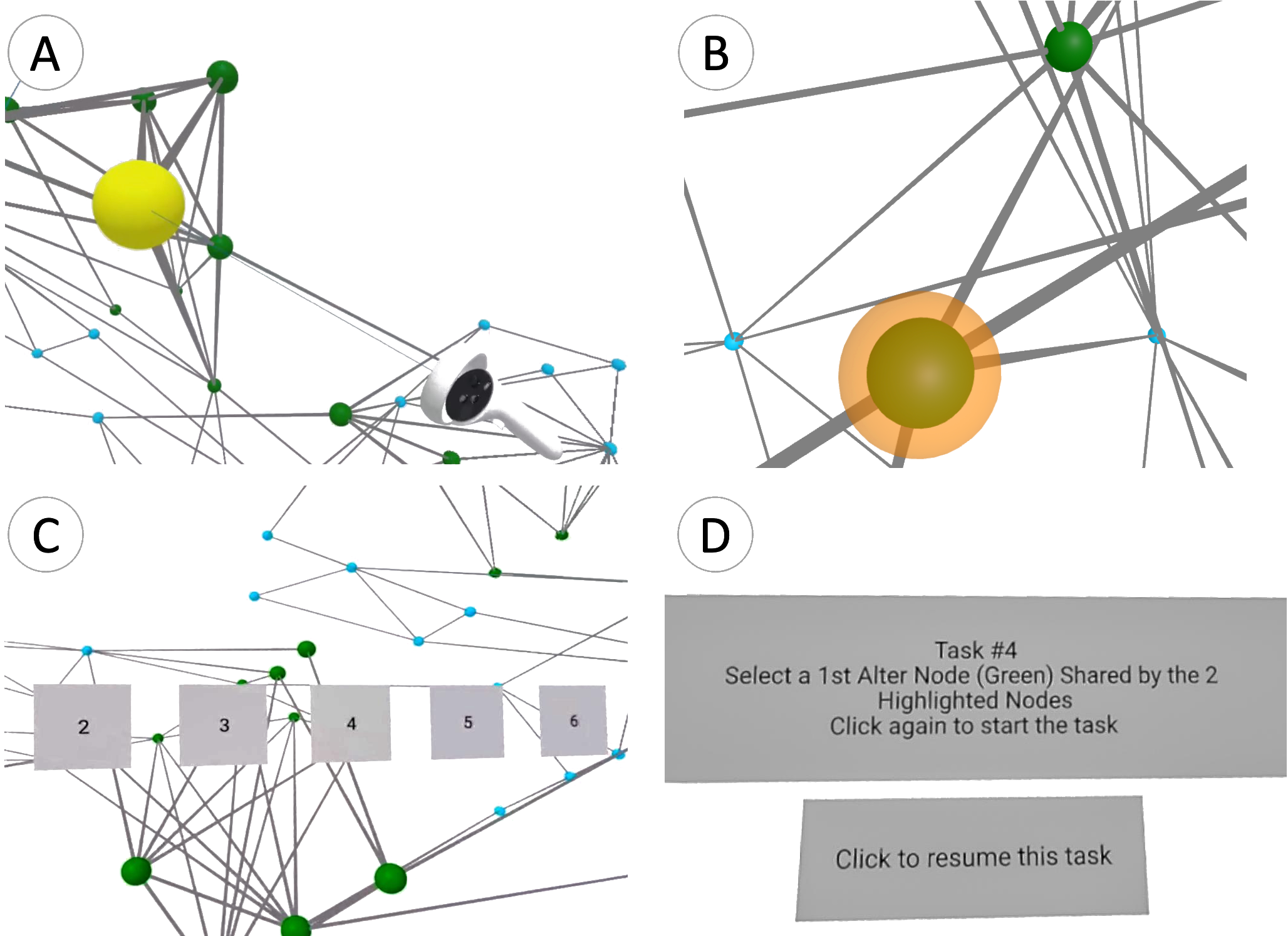}
    \caption{(A) Highlighted nodes by a raycaster. (B) Highlighting of nodes using a halo effect. (C) Response menu for the counting tasks. (D) Result and resume menu.}
    \label{fig:screen-capture}
\end{figure}

\subsubsection*{\circled{P2} \textbf{Task 5: Count 2nd Alters.}}
Participants count the 2nd alters connected to a randomly highlighted 1st alter. 
This task emphasizes identifying nodes that expand the network beyond immediate connections~\cite{wu_egoslider_2016, fu_dyegovis_2021}. 
Unlike Task 3, it focuses on the connections of a single \cc{1st alter}. The highlighted \cc{1st} alter is linked to one to six 2nd alters, a range chosen to \cc{ensure task feasibility.} 
\cc{Participants submit their answers using an interactive panel (\cref{fig:screen-capture}-C).}

\subsubsection*{\circled{P2} \textbf{Task 6: Count 1st Alter Clusters.}}
Participants count the clusters formed by 1st alters, where a cluster is a group of mutually connected nodes. The number of clusters ranges from two to four. This task is critical for understanding the ego's social context and its role in bridging different groups~\cite{ahuja2000collaboration, prell2011social}. Among the topology-related tasks, it requires the broadest view, as participants must observe the largest number of nodes simultaneously.

\subsubsection*{\circled{P1} \circled{P2} \textbf{Task 7: Identify Common Neighbor with Highest Connection Strength to Ego.}}
Participants select the common neighbor with the strongest connection to the ego from among those shared by two highlighted nodes. This task combines Task 1 (strength assessment) and Task 4 (common neighbor identification), requiring simultaneous evaluation of both \cc{connection} strength and network topology. It provides insight into how users integrate \cc{positional and topological cues}. 
As in Task 4, the highlighted node pair shares between three and six common neighbors.

\subsection{Evaluation Metrics}
We assessed participant performance using task completion time, accuracy rate, and movement distance in the immersive space. Completion time was measured from task start to response, which involved clicking a node (Tasks 1, 2, 3, 4, 7) or selecting an answer panel (Tasks 5, 6). \rv{Movement was \cc{recorded} as the total camera travel distance.} 
\cc{We also collected qualitative feedback through a questionnaire and a semi-structured interview.}

\subsection{Experiment Design}
We conducted a within-subjects study comparing the four layouts (\textit{Cube}, \textit{Cylindrical}, \textit{Radial}, and \textit{Spherical}), which span diverse design dimensions.
To minimize learning effects, each participant was assigned a unique network dataset for each layout. 
Since \ac{computing optimal node positions for each dataset} was time-consuming, we used the same dataset across all tasks within each layout, focusing the comparison on layouts rather than tasks. Each participant completed 28 tasks (4 layouts $\times$ 7 tasks).

\subsubsection{Dataset}
We used synthetic datasets derived from a widely studied co-authorship network~\cite{newman2006modularity}, commonly employed in egocentric network analysis. 
To balance realism with experimental control, we \rv{referred} to the structural properties of the larger ego networks in the source dataset, which feature edge densities of 10--30\%, 20--40 first-degree alters, 30--60 second-degree alters, and 2--4 clusters among the 1st alters.
Using these ranges as a reference and confirming task feasibility through pilot testing, \ad{we fixed each dataset at 60 nodes---one ego, 30 1st-degree alters, and 29 2nd-degree alters}---with an edge density of approximately 20\% and two to four clusters among the 1st-degree alters. 
Ego–alter strengths were assigned across ten discrete levels, while all 2nd alters were given a constant, lower strength to reflect their secondary analytical role.

\subsubsection{VR Environment}
The VR environment was built with A-Frame~\cite{aframe}, Three.js~\cite{threejs}, and d3.js~\cite{study_implementation_d3}, and ran in a web browser. Nodes and \cc{links} used Three.js sphere and cylinder geometry with uniform size, \ac{and were placed within a depth range that remained perceptually effective from the initial viewpoint.}
The system ran at 80 FPS on Meta Quest 2 headsets. 
Participants navigated the environment using both physical movement and virtual controls. The left joystick enabled omnidirectional ``flying'' based on headset orientation, while the right joystick supported vertical movement.
\ac{Following A-Frame's default behavior, real-world physical movement was directly reflected in the virtual space at a 1:1 scale via headset position tracking. }
We observed that pilot participants often returned to their starting viewpoint. Therefore, we provided a reset button for quick repositioning.
\ad{We recorded the total camera travel distance as a single combined metric aggregating three components: physical walking, controller-based navigation, and the discrete repositioning produced by each reset. 
Because a reset instantly relocates the camera to the starting viewpoint, it adds travel distance without any corresponding bodily effort. We therefore interpret movement distance as an aggregate indicator of overall navigation activity rather than a measure of physical effort, and complement it with qualitative observations of how participants moved.}
Interaction with nodes was done via raycasters from both controllers. When a ray intersected a node, it turned yellow (\cref{fig:screen-capture}-A), and pressing the trigger confirmed the selection. 
Further details on the implemented layout specifications, including sizes, angles, node dimensions, and movement speeds, are provided in the supplemental materials.

\subsubsection{Participants}
\ac{We recruited 24 participants for the study.} 
\cc{Their VR experience varied: 8 participants had no experience, 10 had low experience, 5 had moderate experience, and 1 had high experience; none reported very high experience.
Familiarity with 3D games, which may relate to spatial navigation, varied as well: 5 participants had no experience, 8 had low experience, 3 had moderate experience, 2 had high experience, and 6 had very high experience. 
Most participants had limited experience with network visualization: 8 had none, 7 had low experience, 4 had moderate experience, 2 had high experience, and 3 had very high experience.}

\subsubsection{Experiment Protocol}
After signing a consent form and completing a demographic questionnaire, participants were introduced to egocentric network concepts (e.g., ego and alters) using printed handouts, which also explained each task with 2D node-link diagrams and the layouts' encoding methods with VR screenshots (see supplemental materials).
Before the main tasks, participants wore the VR headset and explored the immersive environment.
\ac{The tutorial included practice with selecting highlighted nodes, clicking menu windows, and navigating via physical and controller-based movement, following the instructors' guidance. 
For this phase, we used a 3D force-directed layout distinct from those in the main experiment. 
Participants completed the tasks for each layout in an order counterbalanced by a Latin square design to mitigate order effects such as fatigue or learning.}
During the tasks, participants could ask questions, but instructors only referred to information covered in the tutorial to prevent unintended guidance. 
Accidental selections could be canceled and the task resumed (\cref{fig:screen-capture}-D), with completion time and movement data recorded continuously without reset.
After completing all tasks for each layout, participants removed the headset and filled out a questionnaire. Once all four layouts were tested, they ranked their preferences and took part in a brief interview. 
Each session lasted about 80 minutes, and participants received \$10 as compensation.

\subsection{Guiding Questions}
Based on the layout design dimensions, task characteristics, and prior research reviewed in Section~\ref{sec:related_work}, we developed the following guiding questions along with our expectations:

\subsubsection*{\textbf{GQ1: Which layouts support accurate and efficient assessment of ego–alter connection strength, and which hinder it?}}
This addresses property \circled{P1}. \rv{We expect \textit{Radial}, with anchored depth encoding, to facilitate accurate strength assessment.} 
While \textit{Cube} uses non-anchored encoding from the initial viewpoint, it also allows side-view inspection, making it effective. In contrast, \textit{Cylindrical} and \textit{Spherical} may impair depth perception due to the lack of a reference plane, potentially hindering performance.

\subsubsection*{\textbf{GQ2: Which layouts facilitate accurate and efficient identification of alter topology?}}
This addresses property \circled{P2}. Understanding topology requires minimizing node and edge occlusion. We expect the \textit{Spherical}, with minimal spatial constraints and occlusion, to perform well. In contrast, \textit{Radial}, which flattens nodes onto a 2D plane and imposes strict layout constraints, is likely less effective for topological tasks.

\subsubsection*{\textbf{GQ3: Which layout best supports simultaneous understanding of strength and topology?}}
This question explores how layouts perform when tasks require both \circled{P1} and \circled{P2}. Since layouts \cc{may excel in one aspect but not the other}, we aim to identify which layout offers the best balance for integrated tasks.

\subsubsection*{\textbf{GQ4: Which layouts are most and least preferred by users?}}
Beyond task performance, user experience matters. 
\rv{Viewpoint-optimized, exocentric layouts tend to reduce discomfort such as motion sickness; we expect \textit{Cylindrical} to be most preferred.}

\subsubsection*{\textbf{GQ5: How do user strategies differ across layouts?}}
In \textit{Cylindrical} and \textit{Spherical}, users can complete tasks with minimal movement, relying mainly on head motion due to good visibility from the starting point.
In contrast, \textit{Cube} and \textit{Radial} require more movement and frequent viewpoint adjustments to resolve occlusions and interpret positional encoding.

\section{Result}
\label{sec:result}
\begin{figure}[t]
    \centering
    \includegraphics[width=\linewidth]{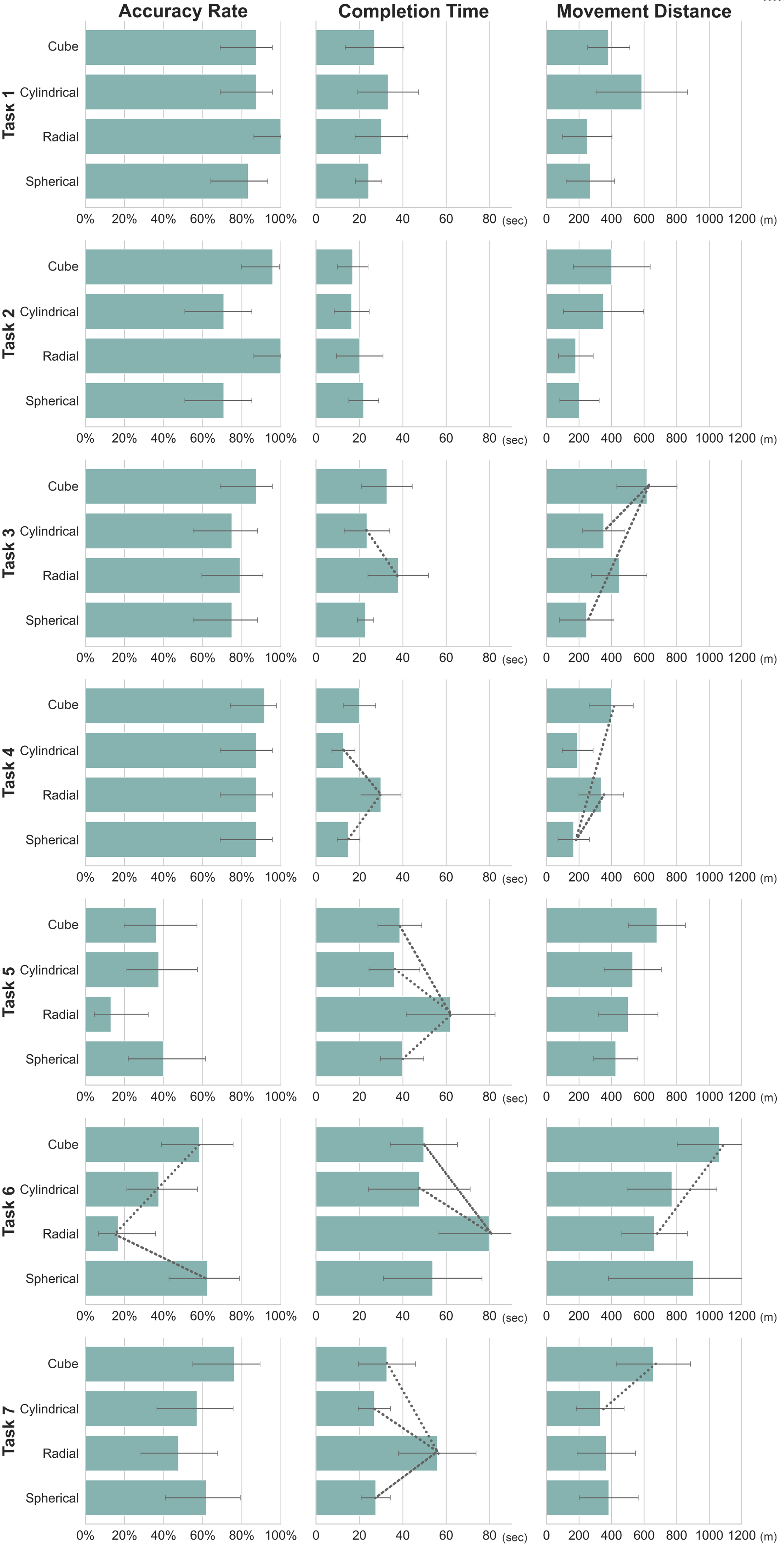}
    \caption{The Results for Mean Accuracy Rate, Completion Time (seconds), and Movement Distance (meters): The confidence intervals indicate a 95\% confidence level for the mean values. Dashed lines signify significance in the pairwise tests.}
    \label{fig:result_task}
\end{figure}

\begin{figure*}[ht]
    \centering
    \includegraphics[width=\linewidth]{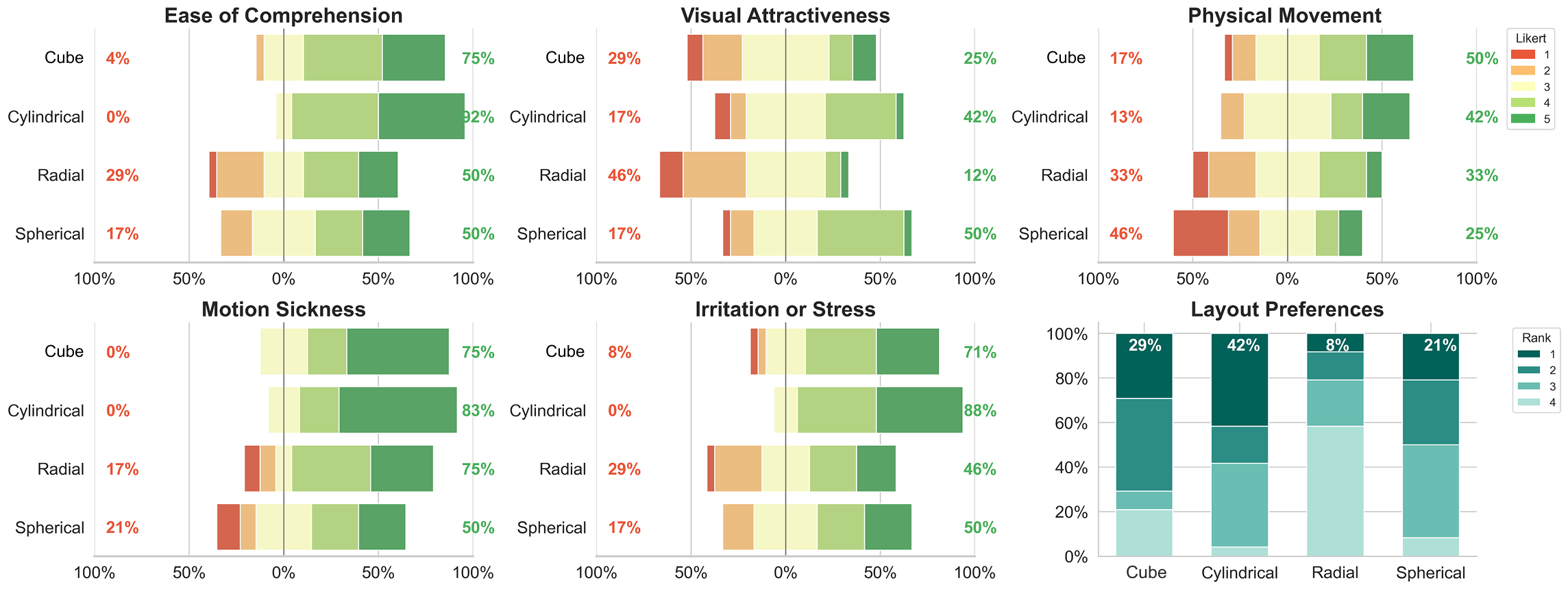}
    \caption{Subjective Feedback Results: Ratings of 1–2 indicate negative responses, and 4–5 positive ones, centered on a neutral rating of 3. Layout preference percentages reflect the proportion of participants selecting each layout as their top choice.}
    \label{fig:result}
\end{figure*}

We analyzed task accuracy, completion time, movement distance, and subjective feedback from the user study.
For each task, we assessed differences among the four layouts with an omnibus test followed by pairwise post-hoc comparisons for the six layout pairs \({4 \choose 2} = 6\). \ad{Because accuracy is a binary outcome in a within-subjects design, we used Cochran's Q test for the omnibus comparison and McNemar's exact test for the pairwise comparisons.}
Completion times and movement distances, being continuous, were analyzed with the Friedman test followed by Wilcoxon signed-rank tests.
All pairwise comparisons used a significance level of $\alpha = 0.05$ with Bonferroni correction ($\alpha = 0.05/6 = 0.0083$).
Subjective ratings (1–5) were aggregated by score frequency, with scales aligned so that 1 indicated the least favorable response (e.g., low visual clarity, high motion sickness) and 5 the most favorable (e.g., clear layout, low physical effort).
Layout preferences were summarized by counting how often each layout was ranked in each position. Task performance results are shown in \cref{fig:result_task}, with 95\% confidence intervals. Subjective ratings appear in \cref{fig:result}.

\subsection{Understanding Ego-Alter Strength} 

\cc{\textbf{Task 1} yielded high accuracy rates for all layouts, with \textit{Radial} performing perfectly (100\% accuracy), aligning with our expectations in \textbf{GQ1}.}
Completion times showed no significant differences. 
However, \textit{Spherical} and \textit{Radial} required less movement. We observed that users could complete the task using head movements and visual scanning from the starting position. In contrast, users in the \textit{Cylindrical} and \textit{Cube} layouts often moved around to inspect nodes from multiple angles.

In \textbf{Task 2}, the accuracy rates were higher for the \textit{Cube} (96\%) and \textit{Radial} (100\%), both of which allow flexible viewpoint adjustment, compared to the more constrained \textit{Cylindrical} and \textit{Spherical} (71\% each). 
\ad{The overall difference in accuracy was significant ($p = .003$), although no pairwise comparison survived Bonferroni correction.}
These findings support \textbf{GQ1}, suggesting that layouts providing an effective viewpoint can \cc{enhance the assessment of connection strength, while those} relying solely on multi-plane depth perception result in lower accuracy. 
Although overall differences in completion time and movement distance were significant, pairwise comparisons did not yield significant layout-level differences. However, among the two most accurate layouts, \textit{Radial} required substantially less movement (\cc{180m}) than \textit{Cube} (\cc{401m}), a gap even more pronounced than in Task 1. 
In \textit{Cube}, with nodes distributed across multiple planes, users had difficulty judging depth and often moved laterally to verify distances.
P20 noted, \textit{``Relying on one perspective is a little bit deceiving. I can kind of double-verify it in that way.''}

\subsection{Understanding Alter Topology}

In \textbf{Tasks 3} and \textbf{4}, all layouts yielded high accuracy, but \textit{Cylindrical} and \textit{Spherical} enabled faster completion and less movement due to reduced occlusions and optimized viewpoints. These results support \textbf{GQ2}, suggesting that these layouts facilitate efficient topology comprehension. Users typically completed the tasks near the starting point, relying on head movements and visual scanning rather than active navigation.

In \textbf{Task 5}, although differences were not statistically significant, the \textit{Spherical} achieved the highest average accuracy, with the \textit{Cylindrical} close behind. Both are 3D layouts with low occlusion from the initial viewpoint, which may have aided perception of the 2nd alters positioned farther from the user. Given the small effect, this offers weak support for \textbf{GQ2}. In contrast, the \textit{Radial} had the lowest accuracy and significantly longer completion times, due to its constrained planar structure limiting topological clarity.

In \textbf{Task 6}, \ad{the overall difference in accuracy was significant ($p = .003$), and pairwise comparisons showed that \textit{Cube} and \textit{Spherical}} yielded significantly higher accuracy than \textit{Radial} ($p < .0083$). Completion times were comparable for \textit{Cube} and \textit{Spherical}, while \textit{Radial} was significantly slower, likely due to its constrained layout, which made cluster structures difficult to perceive.
Interestingly, \textit{Cube} and \textit{Spherical} also showed the highest movement distances (\cc{1062m} and \cc{900m}). While high movement in \textit{Cube} was consistent with earlier tasks, the increase in \textit{Spherical} is notable. As per \textbf{GQ5}, this indicates differing task strategies across layouts: to grasp overall topology, many participants in \textit{Spherical} transitioned from an egocentric to an exocentric viewpoint. As P11 noted, ``\textit{It's possible to view all nodes in the Spherical layout from the initial position, but it was challenging to memorize cluster shapes, and I often lost track of counted clusters when they moved out of sight}.''

\subsection{Understanding Strength and Topology}
\textbf{Task 7} required participants to identify a common neighbor with the highest strength to the ego, combining ego–alter strength and alter topology, addressing \textbf{GQ3} (integrated task).
Accuracy was generally lower than in single-focus tasks (Tasks 1 and 4), indicating increased difficulty, but the \textit{Cube} maintained relatively high accuracy. 
Many users approached the task step by step, first identifying topology from the front view and then shifting to a side view to assess strength, resulting in the largest average movement, significantly more than the other layouts.
In contrast, the \textit{Spherical} and \textit{Cylindrical}, optimized for fixed viewpoints, showed more pronounced accuracy drops, highlighting their limitations for integrated analysis.

\subsection{Subjective Ratings}    
We analyzed participants' subjective ratings, focusing on those who gave positive (4 or 5) or negative (1 or 2) scores (\cref{fig:result}).

\subsubsection*{\textbf{Ease of Comprehension}}
Participants generally found strength encoding via depth intuitive across all layouts. Many understood that closer nodes indicated stronger ties to the ego even before tutorial explanations. P15 remarked, ``\textit{The closer nodes represent closer friends}.''
\rc{\textit{Cylindrical} and \textit{Cube}} received the highest ratings for ease of understanding (\rc{92\% and 75\%} positive, respectively), with participants noting that having all nodes in front made strength relationships easy to perceive.

\subsubsection*{\textbf{Visual Attractiveness}}
\textit{Spherical} and \textit{Cylindrical} were preferred for their visual appeal, largely due to their clean initial viewpoint and minimal visual clutter. Several participants highlighted the immersive experience of being surrounded by nodes in the \textit{Spherical} as a key factor in its appeal.

\subsubsection*{\textbf{Physical Movement and Comfort}}
\textit{Radial} and \textit{Spherical}, both offering 360-degree views, were seen as the most physically demanding, despite the highest movement distances occurring in \textit{Cube}. In \textit{Cube}, participants often walked or sidestepped, which felt natural, whereas in \textit{Radial} and \textit{Spherical} they remained stationary while rotating, causing greater fatigue and motion sickness.
As P6 noted, \textit{``Remaining stationary while rotating was more disorienting and physically uncomfortable than moving around.''}

\subsubsection*{\textbf{Overall Preference}}
\textit{Cylindrical} (42\%) and \textit{Cube} (29\%) were the most preferred layouts, \ac{addressing \textbf{GQ4} (layout preference).}
Participants appreciated \textit{Cylindrical} for its minimal occlusion, wide visibility from the initial viewpoint, and reduced need for body rotation. In contrast, \textit{Radial} was the least preferred, \rv{owing to its frequent rotation requirements and low visual clarity.}
\section{Discussion}
\label{sec:discussion}
Our study reveals how egocentric network layouts affect user performance and experience across key analytical tasks, from which we derive design guidelines. The guidelines are bounded by our study conditions and offered as informed suggestions.

\subsection{Implications from Task Categories}
\subsubsection*{\circled{P1} \textbf{Use Anchored Depth Encoding for Accurate \cc{Ego-Alter Connection} Strength Assessment}}
Encoding connection strength via depth is most effective when nodes share a single reference plane. As demonstrated by the \textit{Radial}, anchored depth encoding enables accurate judgment of ego–alter strength, even for distant nodes.

\subsubsection*{\circled{P1} \textbf{Support Multi-Angle Viewing to Improve Depth Judgment}}
Layouts distributing nodes across multiple planes (e.g., \textit{Cube}) benefit from multi-angle viewing.
While non-anchored depth encoding can impair distance perception, multi-angle observation helps verify spatial relationships and improves accuracy. \ac{Incorporating viewpoint adjustment tools, such as rotation or repositioning widgets, can further support non-anchored depth encoding.}

\subsubsection*{\circled{P2} \textbf{Spherical Layout is Effective for Localized Topology Tasks}}
The \textit{Spherical} layout effectively supports topology tasks through its flexible node distribution and reduced occlusion in 3D, aided by an optimized initial viewpoint. 
It is well suited for exploring local network regions, \cc{such as finding common neighbors or counting 2nd alters.} 
However, its reliance on an egocentric perspective can limit tasks requiring a comprehensive network view. 
\cc{When users needed to count 1st alter clusters,} they had difficulty recalling spatial relationships across different regions, leading to greater movement and longer completion times. 
This aligns with prior research indicating that exocentric views offer better support for memory and spatial understanding in overview tasks~\cite{8836087}. 
\ac{Orientation aids, such as annotations or spatial markers, can mitigate this limitation by helping users maintain spatial awareness.}

\subsubsection*{\circled{P2} \textbf{\rc{Avoid Using Radial Layout for Topology Tasks}}}
The \textit{Radial} layout is generally inadequate for topology tasks due to \rc{its restrictive placement constraints, frequent node occlusion, and edge overlaps.}
These limitations hinder the accurate interpretation of network structures.
For topology-focused analysis we recommend exploring more flexible 3D layouts than \textit{Radial}.

\subsubsection*{\circled{P1}\circled{P2} \textbf{Cube Layout Offers a Balance for Combined Tasks}}
The \textit{Cube} layout offers a good balance between representing ego–alter strength and network topology, leading to higher accuracy in complex, combined tasks.
Users often approach these tasks sequentially, \cc{first examining topology from the front and then strength from side views, }achieving efficient and accurate performance at the cost of considerable movement.
\ac{Designers can account for this by supporting easy viewpoint manipulation or layout transitions to reduce user effort and enhance usability.}

\subsubsection*{\textbf{Participants Prefer Layouts with Less Occlusion and an Exocentric View}}
The \textit{Cylindrical} layout ranked highest in subjective evaluations, praised for its low visual clutter, minimal movement requirements, and ease of grasping the overall structure. The \textit{Cube} layout was also well-received due to its similarly exocentric perspective\cc{, which aligns with} preferences found in previous studies~\cite{8836087, egoVsExoYalong}. 
In contrast, users found the \textit{Spherical} layout to be more physically and cognitively demanding, reporting higher cognitive load and stress, \cc{which contradicts} some earlier findings~\cite{effectOfExplorationMode}.

\subsection{Implications for Design Dimensions}
\subsubsection*{\circled{D1} \textbf{Balance Strength Encoding with Topology Clarity}}
A key finding highlights the trade-off between effective ego-alter connection strength encoding and expressive topology representation. 
While connection strength to nearby alters is generally easy to assess, distant nodes are harder to judge, especially with non-anchored depth encoding that hampers depth perception. \cc{Conversely,} anchored depth encoding improves depth clarity but limits spatial layout flexibility. 
Layouts supporting multi-angle viewing (e.g., \textit{Cube}) help navigate this trade-off, letting users shift perspectives and integrate strength and topology cues.

\subsubsection*{\circled{D2} \textbf{\ac{Consider Fixed Viewpoints When Multi-Angle Viewing is Less Critical}}}
Optimizing the initial viewpoint involves a trade-off between clarity and flexibility.
\cc{Layouts with} fixed viewpoints, such as the \textit{Cylindrical} and \textit{Spherical}, \cc{reduce} occlusion and allow quick grasp of topological structures with minimal movement, suiting topology-focused tasks. However, this advantage comes at the cost of limiting multi-angle exploration. \rc{For tasks requiring confirmation from different perspectives, such as assessing distant nodes' depth or grasping full network topology, we recommend opting for layouts that enable multi-angle exploration.}

\subsubsection*{\circled{D3} \textbf{Leverage 3D Layouts in Immersive Environments}}
Our findings show that 3D layouts support intuitive navigation and reduce occlusion through natural movement, even for VR novices, enabling clearer task-relevant views. 
While 2D layouts in VR simplify depth perception and offer a familiar, desktop-like experience, they \cc{underutilize} 3D space, resulting in lower task accuracy, efficiency, and user preference. 
\rv{For networks comparable to those we studied, we therefore recommend prioritizing 3D layouts in immersive environments. For substantially smaller or sparser networks, where occlusion matters less, simpler 2D layouts may suffice or even be preferable, a trade-off that remains to be verified.}

\subsubsection*{\circled{D4} \textbf{Align View Mode with Task Types}}
Choosing between exocentric and egocentric views depends heavily on the analytical goals.
Exocentric views are generally preferred for comprehensive overviews and reduced motion sickness.
In contrast, egocentric views, as in \textit{Spherical}, arrange nodes around the user, reducing occlusion and enabling rapid, sequential inspection of subnetworks.
However, these benefits must be carefully weighed against limitations like fatigue from stationary rotation and difficulty viewing the entire network simultaneously.
\section{Limitations}
\label{sec:limitation}

\subsubsection*{\textbf{Network Diversity and Dynamics}}
\ad{Though derived from real co-authorship data, our ego-networks were modest in size, narrow in density, and static. 
Since network size affects performance~\cite{feyer_2d_2023}, denser or larger graphs would introduce more occlusion and clutter; \textit{Spherical}, with its spacious arrangement, may better accommodate them, though this remains unverified. Real-world analysis also often involves multiple egocentric networks or temporal dynamics~\cite{spreadline, liu_egocomp_2017, wu_egoslider_2016}.
Future work should test how our findings extend to larger, denser, and dynamic settings.}

\subsubsection*{\textbf{Participant Expertise}}
\ad{Although our tasks mirror common egocentric network analysis tasks that experts perform, our participants were not domain experts. Analysts' exploration strategies and workflows are strongly shaped by domain-specific goals and expectations about network structure~\cite{9903291}; experts may thus prioritize different cues or navigate differently, affecting both performance and layout preferences. The applicability of our findings to professional contexts remains to be verified with experts.}

\subsubsection*{\textbf{Confounding of Design Dimensions}}
\ad{The four layouts differ along multiple design dimensions simultaneously. This holistic comparison evaluates complete, coherent designs as used in practice, but prevents isolating any single dimension's effect; performance differences should therefore be attributed to layouts as a whole rather than to individual design choices.}

\section{Conclusion}
\label{sec:conclusion}
This study identified desirable properties and design dimensions for visualizing egocentric networks in immersive environments. Drawing on these insights, we proposed four novel layouts: \textit{Cube}, \textit{Cylindrical}, \textit{Radial}, and \textit{Spherical}. We evaluated these layouts through a user study with 24 participants performing representative egocentric network analysis tasks. 
While no single layout proved universally optimal, \textit{Cube} was particularly effective for assessing ego–alter connection strength, whereas \textit{Spherical} \rv{was more effective for conveying alter topology}.
The study \cc{also} revealed how layout design shapes users' strategies, affecting movement, task efficiency, and subjective experience.
Based on these results, we derived design guidelines, such as aligning viewpoint optimization with task demands and balancing the clarity of strength encoding against topological expressiveness.
These findings contribute to the growing body of research on immersive network visualization.


\bibliographystyle{src/ACM-Reference-Format}
\bibliography{reference}

\end{document}